\documentclass{webofc}
\usepackage[varg]{txfonts}   
\usepackage{listings}
\usepackage{hyperref}
\begin{document}
\title{Asteroseismology of the {\it Kepler} target KIC\,9204718}
%
%
\author{C. Ulusoy\inst{1}\fnsep\thanks{\href{mailto:cerenulusoy@gau.edu.tr}{\tt cerenulusoy@gau.edu.tr}}, I. Stateva\inst{2}, I.kh.Iliev\inst{2}, B.Ula\c{s}\inst{3}, and M. Napetova\inst{2}}

\institute{School of Aviation, Girne American University, University Drive, PO Box 5, 99428 Karmi Campus, Karao\u{g}lano\u{g}lu, Kyrenia, Turkey
\and
          Institute of Astronomy with NAO, Bulgarian Academy of Sciences,
          blvd.Tsarigradsko chaussee 72, Sofia 1784, Bulgaria
\and
          {I}zmir Turk College Planetarium, 8019/21 sok., No: 22, \.{I}zmir, Turkey
          }
\abstract{%
 The high precision data obtained by the  {\it Kepler} satellite allows us to detect hybrid type pulsator candidates more accurately than the data obtained by ground--based observations. In this study, we present preliminary results on the new analysis of the  {\it Kepler} light curve and high resolution spectroscopic observations of pulsating Am star KIC \,9204718. Our tentative analysis show that the star has hybrid-type pulsational characteristics.
}
\maketitle

\section{Introduction}\label{sec:intro}
New generation space missions provide accurate detections of hybrid--type pulsations in a variety of different stars. 
Recently, using the data of A stars from the {\it Kepler} Telescope,\cite{RefJ1} reported that six Am stars have been known in the  {\it Kepler} field showing $\delta$ Sct type pulsations.The star, KIC\,9204718 (HD\,176843, A3mF0) was first classified as an Am star by \cite{RefJ2}. 
KIC\,9204718 was also listed as a marginal Am type star showing  $\delta$ Sct type pulsations to be a possible candidate of a contact binary system (\cite{RefJ3}).Since Am stars were known as a class of non-pulsating variables this new result makes them very important to test the new diffusion scenario regarding their location in the HR Diagram (\cite{RefJ4}) .

\section{ {\it Kepler} Photometry}\label{sec:sec-1}
The {\it Kepler} time--series of KIC\,9204718  were used to derive pulsational content.The frequencies were extracted from the Long Cadence data by using the software package {\tt SigSpeC} (\cite{RefJ5}). The light curve of KIC\, 9204718  is therefore dominated by two peaks with frequencies $f_{1}$= 0.11421 d$^{-1}$ and $f_{2}$= 0.024384 d$^{-1}$.
\section{Spectroscopic Observations}\label{sec:sec-2}
 Spectroscopic observations have been carried out with ESPERO-a new-commissioned fiber-fed echelle spectrograph attached to the 2-m RCC telescope of Rozhen National Astronomical Observatory. High dispersion spectroscopic data were used to extract the projected rotational velocity and atmospheric parameters.
\section{Conclusion}\label{sec:sec-3}
\begin{figure*}
\centering
\sidecaption
\includegraphics[width=8cm, clip]{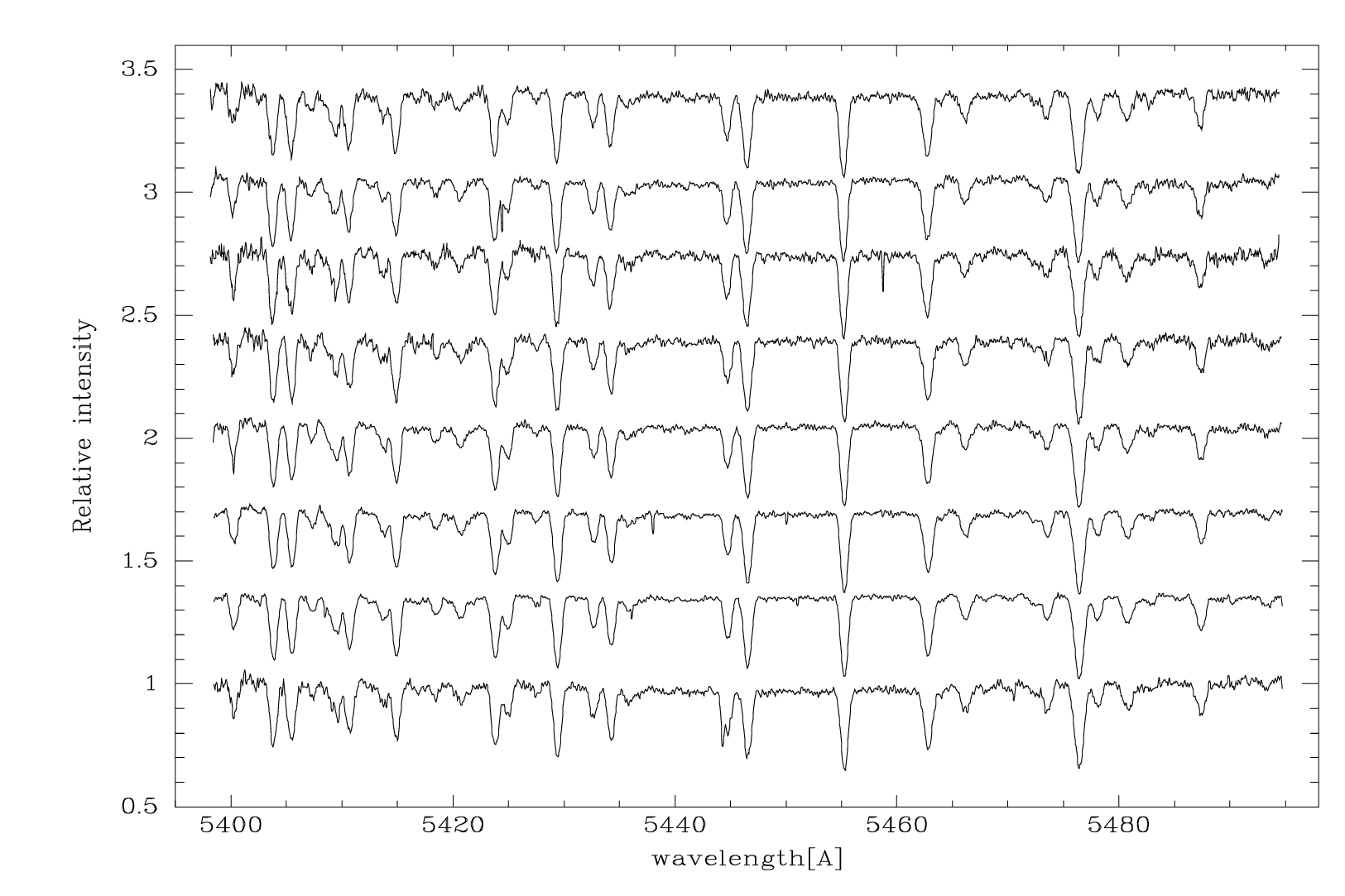}
\caption{Normalized spectra of KIC\,9204718}
\label{fig:fig-1}      
\end{figure*}

We have presented preliminary results of both
frequency and spectrum analysis of marginal Am
type star KIC\, 9204718 observed by the {\it Kepler}
satellite.The radial velocity measurements reveal not very significant variations (Table~\ref{tab:tab-1}). The radial velocity is changed in the framework of 5 km/s. A careful check of the obtained spectra does not show any signs of binarity (Figure~\ref{fig:fig-1}) In conclusion, more observations may enable us to identify at least some of modes and our present work will support in future asteroseismic studies of this star.
 
 \begin{table}
 	\centering
 	\caption{Radial velocity measurements of KIC\, 9204718}
 	\begin{tabular}{llll}
 		\hline
 		Date			&	HJD2457+		&	RV[km/s]&	RVerr[km/s] \\
 		\hline
 		21.07.2016	&	591.44155	& -14.94	&$\pm$ 1.88 \\
 		22.07.2016	&	592.40175 & -15.05	&	$\pm$ 1.82	\\
 		23.07.2016	& 593.44898	& -14.85	&	$\pm$ 1.80	\\
 		17.08.2016	& 618.41679	& -16.16	&	$\pm$ 1.98	\\
 		20.08.2016	& 621.36269 & -16.43	&	$\pm$ 2.29	\\
 		15.09.2016	& 647.48922 & -19.71	&	$\pm$ 1.94	\\
 		16.09.2016	& 648.32971	& -19.65	&	$\pm$ 2.03	\\
 		17.09.2016	& 649.38379 & -19.61	&	$\pm$ 2.11	\\
 		\hline
 	\end{tabular}
 	\label{tab:tab-1}  
 \end{table}
 


%





\begin{acknowledgement} 

\noindent\vskip 0.2cm
\noindent {\em Acknowledgments}: The authors acknowledge the whole  {\it Kepler} team for providing the unprecedented data sets that makes these results possible. CU wishes to thank Miss.Cemile Gür for her valuable support. MN acknowledges for the support of the project DFNP-1 03/11.05.2016 "Support for young scientists" funded by the Bulgarian Academy of Sciences. 
\end{acknowledgement}

%
%

\end{document}